\documentclass{emulateapj} 
\usepackage{latexsym}
\usepackage{mathrsfs} 
\usepackage{times} 
\usepackage{graphics}
\usepackage[german,english,american]{babel}

\newcommand{\sci}{Science}

\newcommand{\scrW}{\mathcal{W}}
\newcommand{\scrM}{\mathcal{M}}

\begin{document}
\shorttitle{The kinetic power of XRB jets}
\title{Estimating the kinetic
  luminosity function of jets from Galactic X-ray Binaries}

\shortauthors{}
\author{
S.~Heinz$^{1,2}$, H.J.~Grimm$^{3}$\\
$^{1}$ Center for Space research, MIT, 77 Mass. Ave., Cambridge, MA 02139\\
$^{2}$ Chandra Fellow\\
$^{3}$ Center for Astrophysics, 60 Garden Street, Cambridge, MA 02138}
\date{23 March 2005}

\begin{abstract}
By combining the recently derived X-ray luminosity function for
Galactic X-ray binaries (XRBs) by Grimm et al.~(2002) and the
radio--X-ray--mass relation of accreting black holes found by Merloni
et al.~(2003), we derive predictions for the radio luminosity function
and radio flux distribution (logN/logS) for XRBs. Based on the
interpretation that the radio--X-ray--mass relation is an expression
of an underlying relation between jet power and nuclear radio
luminosity, we derive the kinetic luminosity function for Galactic
black hole jets, up to a normalization constant in jet power. We
present estimates for this constant on the basis of known ratios of
jet power to core flux for AGN jets and available limits for
individual XRBs. We find that, if XRB jets do indeed fall on the same
radio flux--kinetic power relation as AGN jets, the estimated mean
kinetic luminosity of typical low/hard state jets is of the order of
$\langle W_{\rm XRB}\rangle \sim 2 \times 10^{37}\,{\rm ergs\,s^{-1}}$,
with a total integrated power output of $W_{\rm XRB} \sim 5.5 \times
10^{38}\,{\rm ergs\,s^{-1}}$.  We find that the power carried in
transient jets should be of comparable magnitude to that carried in
low/hard state jets.  Including neutron star systems increases this
estimate to $W_{\rm XRB,tot} \sim 9\times 10^{38}\,{\rm
ergs\,s^{-1}}$. We estimate the total kinetic energy output from
low/hard state jets over the history of the Galaxy to be $E_{\rm XRB}
\sim 7 \times 10^{56}\,{\rm ergs}$.
\end{abstract}
\keywords{}

\section{Introduction}
\label{sec:introduction}
X-ray binaries (XRBs) have long been thought of as classic examples of
accretion flows with little or no complicated outflow physics in the
way of understanding the dynamics of these systems.  It was not until
recently that jets and outflows were realized as important ingredients
in the process of accretion even in XRBs.  The reason for this late
appreciation is the fact that black hole XRBs are less radio loud than
AGNs by about 4 orders of magnitude \citep{heinz:03a}.  

Accreting neutron star binaries, which are now also known to regularly
produce radio emitting jets \citep{fender:05}, are yet another factor
of 30 more radio quiet than black hole XRBs
\citep{fender:01,migliari:03}.  This further reduction in radio flux
has made quantitative analysis of neutron star jets difficult to the
point that rather little is known about them at present.  Therefore,
we will henceforth focus on the properties of jets from black hole
XRBs throughout most of this paper and will present an extension of
our work in \S\ref{sec:neutronstars} to neutron star XRBs.

In consequence, while XRBs are among the brightest X-ray sources in
the sky, their radio output is rather weak.  This, and the fact that
the number of Galactic XRBs is much smaller than the number of AGNs,
are to blame for the fact that very little information about the radio
power from XRB jets exists and that the physical properties of these
jets are even less well known that those of AGN jets (which by
themselves still pose many questions as to their exact makeup and
dynamics).

Two types of radio loud jet-like outflows have been observed in black
hole XRBs.  Following transitions in the X-ray spectral state, one can
often observe bright radio flares, which are optically thin and show a
powerlaw temporal decay on day timescales. The radio emission can be
resolved on sub-arcsecond scales and shows proper motion
\citep{mirabel:94,hjellming:95,fender:99}.  These ejections can carry
a large amount of power, but are relatively rare.  The second kind of
jet-like outflow is associated with the so-called low/hard state in
black hole XRBs \citep[see][for a thorough review on X-ray states in
XRBs]{mcclintock:03}.  These outflows show a flat, optically thick
radio spectrum and typically do not show a temporal decay.  Only two
of these sources have been resolved, showing a collimated jet on
milli-arcsecond scales \citep{stirling:01,dhawan:00b,fuchs:03}.
Consequently, these types of flows are referred to as
``steady/compact'' jets, and for the remainder of this paper, we will
be mostly concerned with these types of jets.

Steady, compact, flat spectrum jets are observed in virtually every
Galactic black hole XRBs in the low/hard state that is accessible to
radio instruments.  Significant progress in our understanding of XRB
jets has recently been made when a relation between the radio
luminosity from these steady, compact jets and the hard X-ray
luminosity was discovered in low/hard state source
\citep{corbel:02,gallo:03} .  This relation has since been found to
extend from stellar mass black holes all the way up the black hole
mass scale to AGN jets \citep{merloni:03,Falcke:04}.  In the remainder
of this paper we will refer to this relation as the ``fundamental
plane of black hole activity'' (FP hereafter).

The FP relation expresses both the fact that the radio luminosity from
compact jets is non-linearly correlated with the X-ray flux, as had
already been discovered in the case of XRBs, as well as the non-linear
dependence of radio flux on black hole mass. The latter is responsible
for the fact that AGN are so much more radio loud than XRBs.  These
relations can be understood naturally if the physics underlying jet
formation (ultimately, strong gravity and MHD in the inner accretion
flow) are invariant under changes in black hole mass
\citep{heinz:03a}.

This scale invariance implies that a relation between the radio
luminosity $L_{\rm r}$ emitted by the jet and the kinetic jet power
$W_{\rm kin}$ is underlying the radio-X-ray-mass relation.  For
reasonable parameters\footnote{For an electron spectrum with powerlaw
index $p=2$ and proportionality between kinetic jet power and magnetic
energy density, $W_{\rm kin}\propto B^2$, i.e., for magnetically
driven jets}, this relation takes on the form $L_{\rm rad} \propto
W_{\rm kin}^{1.42 - \alpha_{\rm r}/3}$, where $\alpha_{\rm r} \approx
0$ is the radio spectral index.  Together with the recent discovery of
the FP, this relation can provide a powerful diagnostic for studying
the properties of jets.

A second recent insight that has spurred heightened interest in the
statistical study of XRBs is the firm determination of the Milky Way
XRB X-ray luminosity function \citep{grimm:02}.  In this paper, we
will couple the information contained in the XRB luminosity function
with the predictive power of the FP relation to derive some of the
missing statistical properties of the XRB jet population needed in
planning observation campaigns for XRBs and in modeling the jets of
XRBs and their impact on the interstellar medium.

In \S\ref{sec:kplf} we will derive the kinetic luminosity function for
Milky Way low/hard state XRB jets, parameterized by the unknown
normalization of the kinetic power -- radio power relation underlying
the fundamental plane relation.  In \S\ref{sec:kinetic}, we will
present an estimate of that normalization derived from AGN jets and
the available limits for individual XRBs, and apply it to the XRB jet
population to derive the absolute kinetic luminosity function.  In
\S\ref{sec:radio} we derive, as corollaries to the kinetic luminosity
function, the predicted radio luminosity function of XRB jets and the
predicted Galactic radio logN/logS distribution.  Section
\ref{sec:summary} summarizes our results.

\section{Deriving the kinetic luminosity function of black hole XRBs
  in the low/hard state}
\label{sec:kplf}
The discovery of the FP relation has enabled a number of diagnostic
tools to be developed for accreting black holes
\citep{merloni:04,maccarone:04,heinz:04a,heinz:04b,markoff:05}.  Since
the current theoretical understanding of the FP relation is based on a
relation between kinetic jet power and radio luminosity, it is now
possible to convert a radio luminosity function into a kinetic
luminosity function \citep{heinz:05b}.  This is a powerful and useful
technique for AGN jets, where we can measure the radio luminosity
function and have good estimates of the kinetic power in a number of
sources. In XRBs we have neither, due to the characteristically lower
radio fluxes of XRB jets (which has delayed their mass discovery by 30
years relative to the large known sample of AGN jets) and the lack of
good estimators of kinetic power in these sources.

However, the small dispersion in black hole mass in black hole XRBs
and the recently published X-ray luminosity function for XRBs
\citep{grimm:02} allow us to use the FP relation to construct both a
predicted radio luminosity function and a kinetic luminosity
function. These estimates are necessarily subject to considerable
uncertainty from a number of unknowns: (a) The conversion of the total
XRB luminosity function to a black hole luminosity function
(essentially, the black hole fraction of XRBs), (b) the normalization
of the kinetic jet power for a given radio power (this holds for AGNs
as well), and (c) the mass distribution of stellar mass black
holes. We will parameterize our results in terms of the fiducial
values we assume for these unknowns to allow for easy adjustment with
future improvements in the knowledge of these parameters.

\subsection{The XRB luminosity function}
The FP relation implies that below a certain critical accretion rate
$\dot{m}_{\rm crit}\equiv \dot{M}_{\rm crit}/\dot{M}_{\rm Edd}\sim
0.01$, there exists a correlation between the mass of a black hole
$M$, its radio luminosity $L_{\rm r}$, and its X-ray luminosity
$L_{\rm x}$ of the form
\begin{eqnarray}
L_{\rm r} & = & L_0\,l_{\rm x}^{0.6} M^{0.78} \label{eq:fp} \\ L_0 &
 \equiv & 1.6\times10^{30}\,{\rm ergs\,s^{-1}}
\end{eqnarray}
where, for convenience, we expressed the X-ray luminosity in terms of
the Eddington luminosity $L_{\rm Edd} = 1.3\times 10^{38}\,{\rm
ergs\,s^{-1}}$ of a one solar mass object: $l_{\rm X} \equiv L_{\rm
X}/L_{\rm Edd}$.  Following the interpretation of
\cite{merloni:03,fender:04}, we will assume in the following that the
transition at $\dot{m}_{\rm crit}$ is due to a state change as
observed in XRBs, where low luminosity sources switch into the
low/hard state.  In XRBs, which we are concerned with here, steady,
flat spectrum jets are closely associated with the low/hard state.
The details of this transition are irrelevant, we will simply use the
associated X-ray luminosity as an upper bound on the luminosity
function.  For a fixed black hole mass $M = 10\,M_{\odot}\,M_{10}$,
this transition corresponds to a critical X-ray luminosity $l_{\rm
X,crit} = 10\,M_{10}\,\dot{m}_{\rm crit}$.
\cite{miyamoto:95,maccarone:03b} showed that the transition from
high/soft to low/hard state actually occurs over a range of values for
$\dot{m}_{\rm crit}$, and that a hysteresis exists with sources on the
descending luminosity branch staying in the high/soft state longer (to
lower luminosities) than ascending sources, which make the state
transition from hard to soft at higher luminosities. \cite{fender:04}
show how the properties of XRB jets correlate with the position of a
source on this hysteresis track.  We will comment on the impact the
uncertainty of $\dot{m}_{\rm crit}$ has on our results in
\S\ref{sec:mdotcrit}.

Following \cite{heinz:03a}, the radio luminosity $L_{\rm r}$ of the
flat spectrum core of the jet is related to the jet power $W_{\rm
jet}$ by
\begin{equation}
W_{\rm jet} = W_{\rm 0}\left(\frac{L_{\rm
r}}{L_0}\right)^{\frac{1}{1.42 - \alpha_{\rm r}/3}}
\label{eq:power}
\end{equation}
where $\alpha_{\rm r}\equiv -\partial(\ln{L_{\nu}})/\partial(\ln{\nu})
\sim 0$ is the radio spectral index an close to zero for the flat
spectrum sources under consideration.  Henceforth we will use
$\alpha_{\rm r}=0$ unless noted otherwise.  $W_{\rm 0}$ is currently
not well known: it is the normalization of the kinetic jet power in
relation to its radio luminosity\footnote{Naively, $W_{0}$ can be
taken as a radiative efficiency, though it should be kept in mind that
most of the radiation is emitted at high frequencies, thus the
radiative efficiency is often times not well defined, as the high
energy cutoff of individual sources is hard to measure and it is not
clear which parts of the spectrum actually arise in the jet and which
arise in the accretion flow.}.  It is not well known because jet power
is still a very difficult quantity to measure, even after four decades
of research on jets.  We will attempt to estimate it below and will
carry it through the algebra as a parameter until then.  Combining
eqs.~(\ref{eq:fp}) and (\ref{eq:power}), we can now write
\begin{equation}
l_{\rm x}=\left(\frac{W_{\rm jet}}{W_0}\right)^{\frac{1.42-\alpha_{\rm
r}/3}{0.6}}M^{-\frac{0.78}{0.6}}
\label{eq:steady}
\end{equation}

We will use the X-ray luminosity function for the Milky Way XRBs
provided by \cite{grimm:02} (see Fig.~12 in their original
paper). Where appropriate, we provide estimates based both on the
actual data they derived (numerically integrated) and parameterized
using the powerlaw approximation $dN/dl=N_0 l^{-\beta}$, fitted by
\cite{grimm:02} to give:
\begin{eqnarray}
\frac{dN_{\rm HMXB}}{dl_{\rm x}} & = & 0.7 l_{\rm x}^{-1.6}
\label{eq:hmxrlf}\\ \frac{dN_{\rm LMXB}}{dl_{\rm x}} & = & 5 l_{\rm
x}^{-1.4}\label{eq:lmxrlf}
\end{eqnarray}
for $l_{\rm x} < 1$ as well as the actual data.  The X-ray luminosity
function does not show a low luminosity cutoff. But at low
luminosities the observed number of sources is very small due to the
sensitivity limits of the ASM on RXTE. However, a comparison with
higher sensitivity (but smaller sky coverage) ASCA data of the
Galactic Ridge Survey \citep{sugizaki:01} show that the XRLF does not
require a cutoff down to $\sim$few $10^{33}$ ergs s$^{-1}$
\citep{grimm:02}.

The radio-X-ray correlation in XRBs has been observed down to X-ray
luminosities of about $l_{\rm x} \gtrsim 10{-4}$ in the case of GX
339-4.  While there is reason to believe that the radio-X-ray relation
might break down at very low luminosities \citep[the synchrotron X-ray
component from the jet scales more slowly with $\dot{m}$, $F_{\rm
x,synch}\propto \dot{m}^{1 - 1.8}$ than the X-rays from the accretion
flow, $F_{\rm x,acc} \propto
\dot{m}^{2-2.3}$][]{merloni:03,heinz:04a,yuan:05}, we will take
$l_{\rm x,min}=10^{-4}$ as a secure upper limit on the low luminosity
cutoff of the radio-X-ray relation.

The relative fraction of black holes as a function of $l_{\rm x}$ in
both the LMXB and HMXB luminosity functions is unknown.  Thus, for
lack of better knowledge, we will use the ratio of observed black hole
to neutron star XRBs, which is about 10\%.  We parameterize this
unknown fraction as $\zeta=[dN_{\rm BH}/dl_{\rm x}]/[dN_{\rm
NS}/dl_{\rm x}]=0.1\zeta_{0.1}$ and assume that the fraction of black
holes is constant as a function of $l_{\rm x}$.  Given the uncertainty
in these estimates and the lack of knowledge of the black hole mass
function, we will henceforth assume that the mean black hole mass is
$10\,M_{\odot}M_{10}$ and evaluate all quantities for this black hole
mass.

The instantaneous kinetic luminosity function for jets from Galactic
black holes is then
\begin{equation}
\frac{dN}{dW_{\rm jet}} = \frac{1.42-\alpha_{\rm r}/3}{0.6
 W_0}M^{-\frac{0.78}{0.6}}\left(\frac{W_{\rm
 jet}}{W_0}\right)^{\frac{0.82-\alpha_{\rm
 r}/3}{0.6}}\frac{dN}{dl_{\rm x}}
\label{eq:dndw}
\end{equation}
for $l_{\rm min}^{0.42} W_0 M^{0.55} < W_{\rm jet} < \dot{m}_{\rm
crit}^{0.42} W_0 M^{0.55}$.  For the powerlaw parameterizations, we
have
\begin{eqnarray}
    \frac{dN_{\rm HMXB}}{dW_{\rm jet}}
    & \approx & \left(\frac{W_{\rm jet}}{W_0}\right)^{\alpha_{\rm r}/3 -
    2.42}\frac{M_{10}^{0.78}}{W_0} \zeta_{0.1}  
    \label{eq:kineticlumi1} \\ 
    \frac{dN_{\rm LMXB}}{dW_{\rm jet}} & \approx & 4 \,\left(\frac{W_{\rm
    jet}}{W_0}\right)^{\alpha_{\rm r}/3 -
    1.95}\frac{M_{10}^{0.52}}{W_0}\zeta_{0.1}
    \label{eq:kineticlumi2}
\end{eqnarray}
in the same bounds.  All that is left to do is the determination of
$W_0$ --- clearly the most difficult and uncertain part of this
exercise.

\section{Estimating the absolute kinetic luminosity function}
\label{sec:kinetic}
\subsection{Estimating $W_0$}
\subsubsection{Estimates of the kinetic power in AGN jets}
Information on the kinetic power from XRB jets is just starting to
become available.  We will discuss the available estimates and how
they can be used to constrain $W_0$ in \S\ref{sec:xrbpower}.  First,
however, we will consider the case of AGN jets, for which information
about the kinetic power is readily available and undisputed in a
number of important cases.  There are three AGN jet sources in the
sample used to derive the FP relation by \cite{merloni:03} which have
reliable estimates of the jet power: M87, Cygnus A, and Perseus A.
These estimates are derived from average powers on very large spatial
scales in the cases of Cygnus A and Perseus A, while for M87 power
estimates exist both for large spatial scales from X-ray cavities, and
for jet scales \citep{bicknell:96}.

We can use these sources to estimate $W_0$ \citep{heinz:05b}. We should
keep in mind, though, that the sources each have some offset from the
fundamental plane relation due to the intrinsic scatter around the
plane. Furthermore, the {\em large scale} power estimates are time
averaged (over times much longer than the dynamical time of the
central jet that is currently producing the radio emission) and might
not be representative of the current jet power.  However, the fact
that the estimate we get from M87 alone (where the large scale power
estimate agrees with the smaller scale estimate) is consistent with
the ones obtained from the other two sources indicates that we are
probably not too far off.  It should also be noted that all of these
sources show compact, flat spectrum jets, which is critical if we want
to use them as templates for low/hard state jets from Galactic XRBs.

The rough power estimates for the three sources are: $W_{\rm jet}
\approx 10^{45.7}\,{\rm ergs\,s^{-1}}$, $10^{44}\,{\rm ergs\,s^{-1}}$,
and $10^{44.5}\,{\rm ergs\,s^{-1}}$ for Cygnus A, M87, and Per A
respectively \citep{carilli:96,bicknell:96,forman:05,fabian:02}, while
their core radio luminosities are $10^{41.4}$, $10^{39.8}$, and
$10^{41.7}\,{\rm ergs\,s^{-1}}$, respectively.  Thus, the
normalization constants we derive for the sources are $W_{0,\rm
CygA}\sim 6.3\times 10^{37}\,{\rm ergs\,s^{-1}}$, $W_{0,\rm M87} \sim
2 \times 10^{37}\,{\rm ergs\,s^{-1}}$, and $W_{0,\rm PerA} \sim
2.6\times 10^{36}\,{\rm ergs\,s^{-1}}$, each carrying {\em
considerable} uncertainty from the estimate of the kinetic power and
from the fact that the sources scatter around the fundamental plane.

We can also calculate what the radio power should be if there were no
scatter around the relation, by estimating the unbeamed radio flux
from the X-ray luminosity and the mass of the objects, using the FP
relation. In this case, the normalization constants are $W_{0,\rm
CygA} \sim 10^{38}\,{\rm ergs\,s^{-1}}$, $W_{0,\rm M87} \sim 6.3\times
10^{37}\,{\rm ergs\,s^{-1}}$, and $W_{0,\rm PerA}\sim 3.7\times
10^{37}\,{\rm ergs\,s^{-1}}$, which is significantly more homogeneous.

This expression is more appropriate to use in an average sense when
considering a large sample of sources, since the radio emission
probably carries the biggest fraction of the scatter in the
fundamental plane relation due to variations in relativistic beaming,
black hole spin, and spectral index.  In this particular case it is
also more appropriate because we are using the X-ray luminosity of
XRBs and the fundamental plane to convert X-ray fluxes into radio
fluxes.  Thus, using X-ray fluxes to calibrate this method seems to us
to be the most consistent approach.  We will therefore use this
``corrected'' estimate of the efficiency, which averages to
\begin{equation}
  W_0 \approx 6.2\times 10^{37}\,{\rm ergs\,s^{-1}} \scrW_{37.8}
  \label{eq:W0}
\end{equation}
with about an order of magnitude uncertainty.  We will carry
$\scrW_{37.8}$ through the rest of the paper to allow the reader to
adjust for future improvements and different personal preferences in
this value.

\subsubsection{Limits from power estimates of XRB jets}
\label{sec:xrbpower}
There is relatively little information about the kinetic power output
of individual Galactic XRBs that we could use to calibrate $W_0$.  The
best information available to date is on the jet in Cygnus X-1.  The
compact jet is resolved on scales of 15 mas, and using a simple
Blandford--Koenigl model \citep{blandford:79} to describe the radio
emission, one can put a lower limit of $W_{\rm jet} > 3\times
10^{33}\,{\rm ergs\,s^{-1}}$, which is very likely much lower than the
true kinetic power.  For the radio flux of 12 mJy, this translates to
a lower limit of $W_0 > 10^{34}\,{\rm ergs\,s^{-1}}$
\citep{stirling:01}.

Based on the IR observations of compact jet in GRS 1915+105,
\cite{fender:00} argue that the normalization for the kinetic power
must be larger than $W_0 > 2\times 10^{35}\,{\rm ergs\,s^{-1}}$,
since, averaged over a sufficient time scale, the kinetic power must
exceed the radiative output of the jet.  Similar arguments are
commonly used in the analysis of Blazar emission \citep{ghisellini:01}.

Detailed modeling of the spectral properties of the XTE J1118 jet has
led \cite{markoff:01} and \cite{yuan:05} to propose values of the
kinetic power that translate to normalization values of $W_0 \approx
10^{38}\,{\rm ergs\,s^{-1}}$ and $2\times 10^{37}\,{\rm
ergs\,s^{-1}}$, respectively.  It is important to note that the jet
power in these models is essentially a free parameter, since
parameters like the proton content and the opening angle of the jets
are unkown.  Nonetheless, these estimates show that reasonable
parametrizations used in the literature for a variety of models are
consistent with the estimate we presented in
eq.~(\ref{eq:W0}). \cite{malzac:05} argue for a lower limit of $W_0
\gtrsim 2\times 10^{38}\,{\rm ergs\,s^{-1}}$ [higher but still
consistent with our estimate in eq.~(\ref{eq:W0}) below] on the basis
of their model for the timing behavior of the accretion flow.
However, this model does not differentiate between a wind-like outflow
and a jet (as long as the flow acts as an energy sink) and it is not
clear how this limit can be turned into a constraint on the jet power
alone.

Calorimetric observations (i.e., radio lobes, X-ray cavities, and
shocks surrounding the cocoons of radio sources) are the most reliable
way to estimate jet powers.  Recently, evidence for the interaction of
the Cygnus X-1 jet with its environment was discovered in the form of
a ring of thermal emission that is probably the shock driven into the
interstellar medium by a so far undetected radio lobe around Cygnus
X-1 \citep{gallo:05}.  The analysis of this shock indicates that the
{\em average} jet power is $10^{36}\,{\rm ergs\,s^{-1}} < \langle W
\rangle < 10^{37}\,{\rm ergs\,s^{-1}}$.  As we will show in
\S\ref{sec:power}, this corresponds to a range of $5\times
10^{36}\,{\rm ergs\,s^{-1}} < W_0 < 5 \times 10^{37}$, which is
consistent with estimate we present in eq.~(\ref{eq:W0}).

While Circinus X-1 does appear to possess a radio lobe, it does not
seem to show a similar shock as Cygnus X-1 does, and thus estimates of
the source age are more difficult.  Furthermore, it is still not
entirely clear whether Circinus X-1 is a neutron star or black hole.
\citep{heinz:02b} estimated the source power to be $\langle W \rangle
\gtrsim 10^{35}\,{\rm ergs\,s^{-1}}$.  This limit is consistent with
that on Cygnus X-1 if the system is in fact a neutron star rather than
a black hole, in which case we would expect the average source power
to be smaller by about the mass ratio, roughly an order of magnitude.

Finally, \citep{kaiser:04} estimate the mean kinetic power of the jets
in GRS 1915+105 from the tentative association with what appear to be
two IRAS hot spots \citep[][note that this association leads to
distance estimates that are in marginally inconsistent with other
measurements]{chaty:01} to be $10^{36}\,{\rm ergs\,s^{-1}} \langle W
\rangle < 10^{37}\,{\rm ergs\,s^{-1}}$.  If a significant potion of
the power in the GRS 1915+105 jet come from the compact/steady flow,
this estimate is consistent with the estimate from above for Cygnus
X-1 and again implies that $5\times 10^{36}\,{\rm ergs\,s^{-1}} < W_0
< 5 \times 10^{37}\,{\rm ergs\,s^{-1}}$, and it is also very much in
line with what we will estimate below based on AGN jets.

\subsection{The absolute kinetic power output of Galactic black holes
  in the low/hard state}
\label{sec:power}
\begin{figure}[tbp]
  \begin{center}
    \resizebox{\columnwidth}{!}{\includegraphics{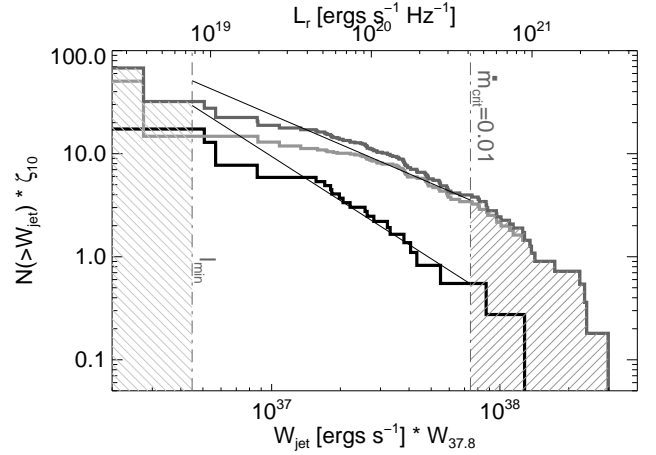}}
  \end{center}
  \caption{Estimate of the cumulative kinetic luminosity function for
  jets from low/hard state black hole XRBs for HMXBs (thin black
  line), LMXBs (thin grey line) and the total (thick grey line), and
  powerlaw approximation for high luminosities.  Also shown (top axis)
  are the corresponding radio luminosities at 5 GHz, for which this
  plot shows the cumulative radio luminosity function.
    \label{fig:kineticlumi}}
\end{figure}

With the estimate of $W_0$ in hand, we can evaluate a number of
important quantities.  The first is simply the kinetic luminosity
function for XRBs.  The approximate analytic expression is given by
eqn.~(\ref{eq:dndw}) with the value of $W_0$ from eq.~(\ref{eq:W0}).
Using the actual data of the XRB luminosity function by
\cite{grimm:02}, we can also plot the kinetic luminosity function, as
shown in Fig.~\ref{fig:kineticlumi}.

The second is the total kinetic power released into the
Galaxy by steady, flat spectrum jets from XRBs.  It is simply the
integral over the luminosity function:
\begin{eqnarray}
W_{\rm tot} & = & \int dW_{\rm jet}\frac{dN}{dW_{\rm jet}} 
  \nonumber \label{eq:totalpower}\\ 
W_{\rm HMXB} & \approx & 2\times 10^{38}\,{\rm
  ergs\,s^{-1}}\,\scrW_{37.8}\zeta_{0.1}
  M_{10}^{0.55}
  \label{eq:totalpowerhmxb}\\ 
W_{\rm LMXB} & \approx &
  3.5\times 10^{38}\,{\rm ergs\,s^{-1}}\,\scrW_{37.8}\zeta_{0.1}
  M_{10}^{0.55}
  \label{eq:totalpowerlmxb}
\end{eqnarray}

Note that the power output from HMXBs is dominated by low luminosity
sources - the total power estimate depends on the assumed lower cutoff
and is thus a lower limit.  For LMXBs, the total power is essentially
a logarithmic function of the lower limit on $W$ to lowest order,
i.e., each decade in $W_{\rm kin}$ contributes the same amount of
total power.  At low luminosities, however, both the HMXB and the LMXB
luminosity functions turn over, the radio-X-ray relation presumably
breaks down, and the total number of XRBs is limited, so, as expected,
the kinetic luminosity function does not diverge.

It is worth pointing out that the value for $W_{\rm tot}$ is somewhat
larger than the estimated total energy output from low/hard state XRB
jets presented in \cite{heinz:02}, which was derived from very
different arguments (we should note that the estimate presented in
this paper should be significantly more reliable).  This also implies
that the statements made in that paper about the total contribution
from XRB jets to the Galactic cosmic ray spectrum still hold (i.e.,
that the total contribution is small, however, spectrally unusual
signatures like narrow lines or Maxwellian features {\em will} stand
out measurably).

Given the estimated star formation rate of $3\,M_{\odot}\,{\rm
yr}^{-1}$ of the Galaxy, and the fact that the HMXB luminosity
function is a good estimator of the star formation rate
\citep{grimm:03}, we can estimate the total kinetic power output in
other galaxies with known star formation rates by multiplying the HMXB
kinetic luminosity estimate from (\ref{eq:totalpowerhmxb}) by the
ratio of star formation rates relative to the Milky Way.  The LMXB
luminosity function does not scale like the star formation
rate. Instead, it should scale like the total mass in stars (dominated
by low mass stars), of which the LMXBs are a given fraction, which is
slowly increasing with time beyond a stellar age of about a billion
years.

Using the stellar mass of the Milky Way, we can also calculate the
time integrated total energy that was released by low/hard state jets:
The total stellar mass in the Milky Way is about
$10^{11}M_{\odot}\scrM_{11}$.  For a current star formation rate of
$\dot{\scrM} \approx 3M_{\odot}\,{\rm yr^{-1}}\dot{\scrM}_{3}$, we can
multiply the current power estimate from HMXBs by the star formation
age of the Milky Way to derive an estimate of the total energy
released by HMXB jets throughout the history of the Galaxy:
\begin{equation}
  E_{\rm HMXB,tot} \approx 2.2\times 10^{56}\,{\rm ergs}\,\scrM_{\rm
  11}/\dot{\scrM}_{\rm 3} \scrW_{37.8} \zeta_{0.1} M_{10}^{0.55}
\end{equation}
while for LMXBs, we should multiply the current energy output by the
age of the LMXB population, which is about $10^{10}\,t_{10}\,{\rm yrs}$
such that
\begin{equation}
  E_{\rm LMXB,tot}=2.3\times 10^{56}\,{\rm ergs}\,t_{10} \scrW_{37.8}
  \zeta_{0.1} M_{10}^{0.55}
\end{equation}

The third quantity worth estimating from this distribution is the mean
jet power of individual XRBs. The main uncertainty here is the
variability of individual sources.  Since the XRLF is a snapshot of
the X-ray output from a large sample of sources, we will have to make
an assumption about how variability of individual sources factors into
the XRLF.  For lack of better knowledge and for reasons of simplicity,
we will consider two possible, simplified scenarios:

\begin{itemize}
\item{The XRLF reflects the temporal X-ray luminosity distribution of
each individual source.  I.e., each source spends a fraction of its
life proportional to $dN/dl_{\rm x}(l_{\rm x})$ in the luminosity bin
$[l_{\rm x},l_{\rm x}+dl_{\rm x}]$.  The luminosity range spanned by
transients is certainly comparable if not larger than the range
spanned by $l_{\rm min}$ and $\dot{m}_{\rm crit}$, supporting this
picture.  Since the XRLF is rather steep, in this scenario an
individual source therefore spends most of its life at low
luminosities - as has long been known in the X-ray community to be the
case for transient sources.

The average kinetic luminosity of an individual source is simply the
total kinetic luminosity of all XRBs divided by the number of black
hole XRBs in the luminosity interval $10^{-4} < l_{\rm x} <
10\dot{m}_{\rm crit}M_{10}$:
\begin{equation}
\langle W\rangle_{\rm HMXB}\approx 1.7 \times 10^{37}\,{\rm
  ergs\,s^{-1}}\,\scrW_{37.8} M_{10}^{0.55}
\label{eq:meanp}
\end{equation}
and
\begin{eqnarray}
\langle W\rangle_{\rm LMXB} & \approx & 2.8 \times 10^{37}\,{\rm
ergs\,s^{-1}}\scrW_{37.8} M_{10}^{0.55}
\label{eq:meanplmxb}
\end{eqnarray}

These are very large numbers indeed, about 1-2\% of the Eddington
luminosity for a 10 solar mass black hole, {\em if} our estimate of
$\scrW_{37.8}$ is at least of the right order of magnitude.  This
indicates that the kinetic energy transported by XRB jets is
comparable to or larger than the radiative output: the integral over
the luminosity functions in eqs.~(\ref{eq:hmxrlf}) and
(\ref{eq:lmxrlf}) gives average X-ray luminosities of $\langle L_{\rm
x,HMXB} \rangle \sim 7 \times 10^{35}\,{\rm ergs}$ and $\langle L_{\rm
x,HMXB} \rangle \sim 1.2 \times 10^{37}\,{\rm ergs}$.  It also implies
that the kinetic energy released in low luminosity states is at least
comparable to the energy advected into the black hole \citep[see
also][]{fender:03}.

We can now compare these numbers with the limits from Cygnus X-1 and GRS
1915+105 from the previous section: $\langle W\rangle_{\rm Cyg X-1} >
10^{36}\,{\rm ergs\,s^{-1}}$ and $10^{36}\,{\rm ergs\,s^{-1}} \langle
W\rangle_{\rm 1915} < few \times 10^{37}\,{\rm ergs\,s^{-1}}$, which
imply $0.1 < \scrW_{37.8} < few$, very much consistent with the
estimate derived from AGN jets.

In this scenario, all sources in either the HMXB or LMXB class are
assumed to be very similar, which, given the diversity of X-ray
binaries, is clearly a simplification.  The fact that the XRLF is not
altered by the variability of individual source \citep{grimm:05}
somewhat supports this simplified picture and the assumption that we
can average over the luminosity function to remove the effects of
variability.}

\item{Alternatively, we could consider a scenario where each source
varies around an interval in $W_{\rm kin}$ much narrower than the
width spanned by $l_{\rm min}$ and $\dot{m}_{\rm crit}$.  In this
case, individual sources differ from each other in average kinetic
power, and the distribution of $\langle W\rangle$ is identical to the
distribution of instantaneous power $W$.  Still, the {\em ensemble
averaged} kinetic luminosity function $\Phi(\langle W_{\rm
jet}\rangle)$ is identical to the snapshot kinetic luminosity function
$\Phi(W_{\rm kin})$ and the number derived for the total kinetic
power is identical to eq.~(\ref{eq:totalpower}).}
\end{itemize}

Neither of these scenarios is likely to be entirely correct; however,
the range in $W_{\rm kin}$ spanned by the kinetic luminosity function
in eq.~(\ref{eq:dndw}) is relatively narrow and it is rather likely
that individual sources traverse a range in $W_{\rm jet}$ that is at
least as large as this.  Thus, using the average value from
eq.~(\ref{eq:meanp}) for individual sources to estimate the effects on
the interstellar medium appears to us to be a reasonable choice.

Finally we stress again that these estimates are based on an
extrapolation from AGN jet power estimates, which themselves are
somewhat uncertain and are thus working estimates only.  Obviously, a
much better and more accurate way to estimate $W_{\rm kin}$ for XRBs
is a direct determination.  We anticipate that conclusive measurements
of $W_0$ from XRB radio lobes \citep{heinz:02b} are going to become
available in the near future.

\subsection{Hysteresis and the critical accretion rate}
\label{sec:mdotcrit}
As hinted at in \S\ref{sec:introduction}, the exact location of
$\dot{m}_{\rm crit}$ is not only uncertain, sources actually show a
hysteresis in transitioning from the high/soft state to the low/hard
state and back.  The transition from high/soft to low/hard typically
occurs at lower luminosities than the reverse transition.  Before
elaborating as to how much this will affects our estimates, it is
worth examining quickly what the effects of different values for
$\dot{m}_{\rm crit}$ on $\langle W\rangle$ and $W_{\rm tot}$ actually
are.

We have evaluated eqs.~(\ref{eq:totalpowerhmxb}) and
(\ref{eq:totalpowerlmxb}) for a range of values of $\dot{m}_{\rm
crit}$ and fitted the results with a quadratic in log-log space.  For
the total jet power, we find that
\begin{eqnarray}
  \log{W}_{\rm HMXB} & \approx & 38.34 + 0.05\left[\log(\dot{m}_{\rm
  crit}) + 2\right] \label{eq:mdothmxb}\\ & & -
  0.238\left[\log({m}_{\rm crit}) + 2\right]^2 +
  \log{\scrW_{37.8}\zeta_{0.1}M_{10}^{0.55}} \nonumber \\ \log{W}_{\rm
  LMXB} & \approx & 38.58 + 0.38\left[\log(\dot{m}_{\rm crit}) +
  2\right]
  \label{eq:mdotlmxb}\\ & & - 0.151\left[\log({m}_{\rm crit}) +
  2\right]^2 + \log{\scrW_{37.8}\zeta_{0.1}M_{10}^{0.55}} \nonumber
\end{eqnarray}
Since the mean jet power is simply the total power divided by the
number of binaries, it has the same dependence on $\dot{m}_{\rm
crit}$, but zero order different normalizations, as given by
eqs.~(\ref{eq:meanp}) and (\ref{eq:meanplmxb}).

This demonstrates that our estimates are not very strongly dependent
on $\dot{m}_{\rm crit}$ for values of $\dot{m}_{\rm crit} \gtrsim
0.01$.  For LMXBs, the power changes by a factor of 1.7 when
increasing $\dot{m}_{\rm crit}$ from 0.01 to 0.1.  Given that the
expected range in which this hysteresis occurs is about $0.01 <
\dot{m}_{\rm crit} < 0.1$ \citep{fender:04}, we can estimate the
effect by assuming that sources spend an equal amount of time on both
the high/soft and the low/hard branches.  Thus, a crude estimate would
imply that about 50\% of the sources above $\dot{m}_{\rm crit} > 0.01$
are still in the low/hard state and produce jets. This would increase
our estimates of the jet power for LMXBs by 36\%, well within the
uncertainties of our estimates.  HMXBs are essentially unaffected by
this change because there are very few sources above $\dot{m}=0.01$.

\subsection{Transient Jets}
As mentioned in the introduction, some XRBs display transient jet
emission associated with rapid changes in their X-ray luminosity.
These events can reach radio fluxes that are much larger than in the
case of steady, compact jets produced in the low/hard state.  The
classic example of this kind of source is GRS1915+105
\citep{mirabel:94,fender:99}, the complex behavior of which is too
diverse to be reviewed in the scope of this paper \citep[see][for a
review of GRS 1915+105]{fender:04b}.  Other transient sources for
which this behavior has been observed include GRO J1655-40
\citep{hjellming:95}, V4641 Sgr \citep{orosz:01}, XTE J1748-288
\citep{fender:01}, and XTE J1550- \citep{corbel:02} \citep[see table 1
in][for more details]{fender:05}. The radio emission from these events
is typically optically thin and can be spatially resolved.

In several cases, superluminal proper motion has been detected
\citep{mirabel:94,hjellming:95,fender:99,corbel:02,fender:04c}, which
implies relativistic velocities with Lorentz factors $\Gamma\approx
2-5$.  The relative abundance of measurable parameters in the
optically thin, partially resolved case makes it possible to estimate
the total energy contained in the emission region and estimating the
kinetic power in these transient jets.  We shall briefly discuss these
estimates and their implication for the total jet power from XRBs.

It is currently not understood what the relationship between the
transient jet events observed during state transitions, and the steady
compact jets observed in the low/hard state is.  It is important to
stress that we cannot use the power estimates from the transient
events to normalize the low/hard state kinetic luminosity function.
Because we have limited the kinetic luminosity function derived above
to the luminosity range typically spanned by the low/hard state, our
kinetic luminosity function and the quantities derived from it don't
account for the separate component contributed by these transient jets
by construction.

There are several reasons why we cannot simply expand our treatment to
include these sources: (a) By their nature, these events are
transient, and unlike in the low/hard state, a source may or may not
be emitting a transient jet at a given X-ray luminosity in a given
state.  The fraction of time a transient source is emitting a
transient jet at a given X-ray luminosity is unknown. (b) The
radio-X-ray relation that we used to derive the relation between the
X-ray luminosity and the radio luminosity does not hold for transient
jets. (c) The emission from transient jets is optically thin, and thus
the relation between radio emission and kinetic power in
eq.~(\ref{eq:power}) does not hold (which is one of the reasons why
the radio-X-ray relation breaks down).

In the absence of statistical information about the relative duration,
brightness, and kinetic power of individual transient jet events
(which would be necessary to calculate the integrated transient jet
power), we can still make an educated, though very rough guess of the
contribution of these jets to the total kinetic power: For one thing,
\citep{fender:05} show that sources undergoing a hard--soft transition
are known to produce transient jets, while they speculate that sources
making a soft-hard transition do not.  Thus, over some luminosity
range at the transition luminosity from hard to soft state, we should
expect a transient jet to be emitted in between 50\% to 100\% of the
sources.

Secondly, \cite{fender:05} show that a relation between X-ray power
and kientic power exists in transient jet sources and can be (roughly)
calibrated to give
\begin{equation}
  W_{\rm jet}=1.3 L_{\rm Edd} l_{\rm x}^{0.5\pm 0.2}
\label{eq:transient}
\end{equation}
This relation is surprisingly close to the relation we used above for
the low/hard state jets: $W_{\rm jet}=1.8 L_{\rm Edd} l_{\rm x}^{0.42}
W_{37.8} M_{10}^{0.55}$, as already pointed out by \cite{fender:05}.

Thus, if we knew the duty cycle $\xi$ of transient events (i.e., the
fraction of source above $\dot{m}_{\rm crit}$ which are producing
transient jets), we could graft these two functions together at
$\dot{m}_{\rm crit}$ to make one continuous kinetic luminosity
function and integrate to get the total power.  Since we have neither
a functional form of the distribution of $\dot{m}_{\rm crit}$ nor the
duty cycle of transient events above $\dot{m}_{\rm crit}$, we can only
state that the expression in eq.~(\ref{eq:transient}) is statistically
identical to eq.~(\ref{eq:steady}) and that the contribution from
transient jets to the kinetic power should therefore be well described
by eqs.~(\ref{eq:mdothmxb}) and (\ref{eq:mdotlmxb}) multiplied by
$\xi$, which describe the dependence of the kinetic power on the value
of $\dot{m}_{\rm crit}$.

As already pointed out in \S\ref{sec:mdotcrit}, increasing
$\dot{m}_{\rm crit}$ by an order of magnitude only increases the
kinetic power estimates by a factor of $\approx 1.7$.  Since the
transient jet source fall on essentially the same kinetic luminosity
function with an unknown duty cycle $\xi < 1$, the ratio of the
kinetic power carried in transient jets to that carried in compact,
steady jets should be roughly $0.7\xi/W_{37.8}$.  This gives a total
power estimate of order $4\xi \times 10^{38}\,{\rm ergs\,s^{-1}}$ for
transient jets.

On the other hand, we know from the observed number of outbursts that
the total kinetic power from transient jets in the Galaxy is of the
order of $few \times 10^{38}\,{\rm ergs\,s^{-1}}$ \citep{heinz:02},
compared to the total estimated power of $\approx 5\times
10^{38}\,{\rm ergs\,s^{-1}}$.  Thus, based on the estimates of $W_{0}$
we provided above and the kinetic power normalization of transient
jets by \cite{fender:04}, the integrated kinetic powers of transient
jets and of compact steady jets are of the same order of magnitude and
the duty cycle of transient jets should not be much smaller than
$\approx 10\%$.

\subsection{Neutron Stars}
\label{sec:neutronstars}
As mentioned in \S\ref{sec:introduction}, neutron stars are even less
radio loud than black hole XRBs at typical X-ray luminosities.  At the
current time it is not entirely clear whether the FP relation of
eq.~(\ref{eq:fp}) is applicable to neutron star systems as well: it is
possible that a radio-X-ray relation exists for neutron stars, but has
a different slope \citep{migliari:03}, it is also possible that no
such relation exists at all.  In either case, an extension of our
method is not easily possible.  For simplicity, however, we will
assume that the same correlation shown in eq.~(\ref{eq:fp}) holds for
neutron stars as well, but with a different normalization.

Given the difference in radio loudness of a factor of $\sim 30$ at a
fixed X-ray luminosity and taking into account the mass difference
between neutron stars and black holes of about $M_{\rm BH}/M_{\rm NS}
\sim 10/1.4$, this changes the normalization of eq.~(\ref{eq:fp}):
\begin{equation}
  L_{\rm r} =
  \frac{1}{30}\left(\frac{10}{1.4}\right)^{0.78}L_{0}l_{\rm
  x}^{0.6}M^{0.78}
\end{equation}
Consequently, the normalization of
eqs.~(\ref{eq:kineticlumi1}-\ref{eq:kineticlumi2}) changes as well:
\begin{equation}
  \frac{dN_{\rm NS}}{dW_{\rm jet}} =
      \left[30\left(\frac{1.4}{10}\right)^{\frac{0.78}{1.42-\alpha_{\rm
      r}/3}}\right]^{-\beta}\zeta^{-1}\frac{dN_{\rm BH}}{dW_{\rm jet}}
\end{equation}
The numerical evaluation for $\beta=1.6$ in the case of HMXBs gives
only a slight change of 1.1, while $\beta=1.4$ in the case of LMXBs
gives 1.4. Note, however, that both $\zeta$ and the mass are going to
be different for neutron stars.  Taking these into account, the
absolute normalization the kinetic luminosity function is increased by
factors of $2.4$ and $3.0$ for neutron star HMXB and LMXB
respectively, compared to black hole HMXBs and LMXBs.

\cite{fender:05} suggest that high-field neutron star systems do not
produce jets, as they show no radio emission.  Since a large fraction
of the neutron star HMXBs are, in fact, X-ray pulsars \citep{liu:00}
with associated high fields, the normalization of the kinetic
luminosity function for HMXB neutron stars is likely smaller by at
least a factor of 2 compared to the above estimate.  Furthermore, all
HMXBs are believed to have rather strong fields ($B\gtrsim
10^{12}\,{\rm G}$), which \cite{fender:05} conjectured as being unable
to produce jets at all.  We therefore conservatively assume that the
HMXB values below are upper limits.

Taking into account the lower X-ray luminosity that corresponds to
$\dot{m}_{\rm crit} = 0.01$ in neutron stars, the estimates for the
mean kinetic power from neutron star jets is $\langle W\rangle_{\rm
HMXB} < 2.6 \times 10^{37}\,{\rm ergs\,s^{-1}}\scrW_{37.8}$ and
$\langle W\rangle_{\rm LMXB} \sim 2.5 \times 10^{37}\,{\rm
ergs\,s^{-1}}\scrW_{37.8}$.  We can then estimate the total kinetic
power from neutron stars into the Galaxy to be $W_{\rm LMXB} < 3.4
\times 10^{38}\,{\rm ergs\,s^{-1}}\scrW_{37.8}$ for HMXBs and $W_{\rm
LMXB} \sim 3.3\times 10^{38}\,{\rm ergs\,s^{-1}}\scrW_{37.8}$ for
LMXBs.  This brings the total kinetic power output from XRBs (black
holes and neutron stars) to $W_{\rm XRB} \sim 9\times 10^{38}\,{\rm
ergs\,s^{-1}}$.  Finally, the total energy released by neutron star
LMXB jets over the lifetime of the Galaxy is $E_{\rm LMXB} \approx
2.2\times 10^{56}\,{\rm ergs}\scrW_{37.8} t_{10}$.

\section{A corollary: the radio luminosity function and radio
  logN-logS of XRBs}
\label{sec:radio}
Given the X-ray luminosity function for Galactic XRBs in
eqs.~(\ref{eq:hmxrlf}) and (\ref{eq:lmxrlf}) and the FP relation from
eq.~(\ref{eq:fp}), we can easily write down the predicted Galactic
radio luminosity function for this population of binaries:
\begin{equation}
  \frac{dN}{dl_{\rm r}} = \frac{dN}{dl_{\rm x}}\frac{dl_{\rm
  x}}{d_{\rm r}}
\label{eq:xraylumi}
\end{equation}
where, following the nomenclature from above, $l_{\rm r}=L_{\rm
r}/L_{\rm Edd}$ is the radio luminosity in units of the Eddington
luminosity of a one solar mass object.  The top axis in
Fig.~\ref{fig:kineticlumi} shows the predicted cumulative radio
luminosity function.

For Galaxies with stellar populations of a similar age to the Milky
way, the LMXB distribution should just be proportional to the mass
$M$, while the HMXB distribution should be proportional to the star
formation rate $\dot{M}$:
\begin{eqnarray}
  \frac{dN_{\rm HMXB}}{dl_{\rm r}} & = & 10^{-8.7}l_{\rm
  r}^{-2}M_{10}^{0.78}\zeta_{0.1} \label{eq:hmrlf} \\ \frac{dN_{\rm
  LMXB}}{dl_{\rm r}} & = & 10^{-5.2}l_{\rm
  r}^{-5/3}M_{10}^{0.52}\zeta_{0.1}
\end{eqnarray}
for $10^{-9} < l_{\rm r} < 3.2 \times 10^{-7}\dot{m}_{\rm
crit}^{0.6}$.  Again, while the total emitted radio power in the HMXB
distribution seems to diverge logarithmically if the lower limit
$l_{\rm r,min}$ goes to zero, the luminosity function is, of course,
limited by the number of sources, which is not infinite.  Furthermore,
the radio spectrum will eventually become optically thin at low
luminosities and the luminosity function will become flatter than the
expression in eq.~(\ref{eq:xraylumi}) and converge.  Note that in
X-ray binaries this transition happens at very low luminosities: For
typical low/hard state sources, the break from optically thick to thin
occurs in the infra-red, and the break frequency $\nu_{\rm b}$ roughly
follows $\nu_{\rm b} \propto L_{\rm r}^{8/17}$ \citep{heinz:03a}.  For
the break to move below radio frequencies, we would have to consider
radio fluxes 8 orders of magnitude below what is typically observed.

Finally, we can estimate the flux distribution from the radio
luminosity function (log N - log S).  The flat spectrum XRB radio flux
distribution for a galaxy at distance $D$ is simply $d\log{N}/d\log{S}
= 4\pi D^2 d\log{N}/d\log{l_{\rm r}}$, since all sources are
essentially at the same distance.  Unfortunately, it is clear from the
range in luminosities implied by Fig.~\ref{fig:kineticlumi} that
sources in Galaxies further than about a Mpc will not be observable
any time soon unless the population is large enough to contain
significantly beamed sources (note that the verdict on how
relativistic XRB jets actually are is still out - see, for example,
the discussion in \cite{gallo:03,heinz:04b,kaiser:04,narayan:05}).

To derive the predicted radio flux distribution in the Milky Way, we
have to convolve the luminosity function with the space distribution
of Binaries inside the Milky Way.  Following \citep{grimm:02} and
taking $r$ and $z$ to be the radial and vertical distance to the
Galactic center, the disk populations is describe by the function
\begin{equation}
  n_{\rm disk}(r,z)=\frac{e^{(-\frac{r_{\rm m}}{r}-\frac{r}{r_{\rm
  d}}-\frac{\left|z\right|}{r_{\rm z}})}{8\pi\,r_{\rm m}r_{\rm
  d}r_{\rm z}}}{K_{2}(2\sqrt{\frac{r_{\rm m}}{r_{\rm d}}})}
\end{equation}
with $r_{\rm m} \approx 4\,{\rm kpc},\ r_{\rm d} \approx 3.5\,{\rm
kpc},\ r_{\rm z} \approx 410\,{\rm pc}$ for LMXBs and $r_{\rm m}
\approx 6.5\,{\rm kpc},\ r_{\rm d} \approx 3.5\,{\rm kpc},\ r_{\rm z}
\approx 150\,{\rm pc}$ for HMXBs, and $K_2$ is BesselK.  The LMXB
distribution also requires a spherical halo component, which we will
assume to follow
\begin{equation}
   n_{\rm halo}(R)=\frac{b^{8.5}e^{-b\left(\frac{R}{R_{\rm
   e}}\right)}}{16\pi\,R_{\rm e}^3\Gamma(8.5)}\left(\frac{R}{R_{\rm
   e}}\right)^{7/8}
\end{equation}
with $b\approx 7.7$ and $R_{\rm e}\approx 2.8\,{\rm kpc}$.  Finally,
about 30\% of the disk sources reside in the Galactic center
population, at about 8 kpc distance, for which we use the distribution
\begin{equation}
  n_{\rm bulge}(R)=\frac{(R/R_0)^{-1.8}e^{-(R/R_{\rm
  t})^2}}{2\pi\,\Gamma[0.4]\,R_0^{1.8}\,R_t^{1.2}}
\end{equation}
with $R_0=1.0{\rm \,kpc}$ and $R_t=1.9\,{\rm kpc}$.

\begin{figure}[t]
  \vspace*{18pt}
  \begin{center}
    \resizebox{\columnwidth}{!}{\includegraphics{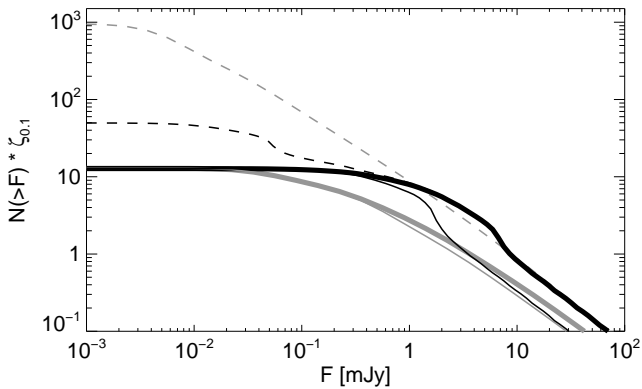}}
  \end{center}
  \caption{Estimate of the radio flux distribution of Milky Way X-ray
  binaries observed at solar radius. Grey: HMXBs, black: LMXBs. Thick
  solid lines: $l_{\rm min} = 10^{-4}, \dot{m}_{\rm crit}=0.01$, thin
  dashed lines: $l_{\rm min} = 0, \dot{m}_{\rm crit}=0.01$, thin solid
  lines $l_{\rm min} = 10^{-4}, \dot{m}_{\rm crit}=0.001$.
    \label{fig:lognlogs}}
\end{figure}

The fraction of LMXBs in the halo is approximately 25\% (including
Globular Cluster sources). We will take the functional form of the
luminosity function to be the same for the halo, bulge, and disk
components.  We then integrate over the luminosity function plotted in
Fig.~\ref{fig:kineticlumi} and arrive at the flux distribution shown
in Fig.~\ref{fig:lognlogs} for different values of upper and lower
cutoff $l_{\rm min}$ and $\dot{m}_{\rm crit}$.  The thick lines
indicate the fiducial luminosity interval of $10^{-4} < l_{\rm x} <
10^{-1}$.  Note that these are time averaged curves.  Temporal
variability necessarily introduces large uncertainty at the high flux
end, where few sources contribute at a given time.

Since we know at least one HMXB source which emits regularly at 10
mJy levels (Cygnus X-1) and a spectrum of other transient sources that
regularly pass above the 10mJy line, the HMXB curve derived for the
fiducial parameters can safely be regarded as a lower limit.

Clearly, the question of how far the X-ray luminosity function extends
to lower luminosities and where the radio-X-ray relation breaks down
has large implications for the number count of sources at lower
fluxes.  In other words, extending the sensitivity of XRB monitoring
campaigns to lower fluxes will reveal rather quickly below which radio
luminosity the predicted radio luminosity function breaks down.

Finally, it should be noted that this is essentially the probability
distribution of finding a given binary at a given distance from earth,
convolved with the predicted radio luminosity function.  Since the
actual number of XRBs is small, the error bars on this curve
especially for large fluxes are probably large.  However, a realistic
assessment of the uncertainties in this distribution is beyond the
scope of this paper.

\section{Conclusions}
\label{sec:summary}
Starting from the observed radio-X-ray-mass relation for accreting
black holes and the observed X-ray luminosity function for XRBs, we
derived the predicted kinetic luminosity function of compact jets from
Galactic black holes in the low/hard state.  The integration over the
kinetic luminosity function of compact jets yields estimates of the
average kinetic power output of $\langle W_{\rm jet} \rangle \sim 2
\times 10^{37}\,{\rm ergs\,s^{-1}}$ (dominated by LMXBs) and total
integrated kinetic energy input over the history of the Galaxy of
$E_{\rm tot} \sim 4.5\times 10^{56}\,{\rm ergs}$.  We argue that
transient jets should carry a comparable amount of power.  We also
derived the predicted radio luminosity function and the radio flux
distribution for XRB jets.  These estimates can be used in future
modeling of XRB jet parameters and provide a base line for estimating
the impact of XRB jets in the interstellar medium.

\acknowledgements{After submission of this manuscript we became aware
  of a related paper \citep{fender:05b} recently accepted for
  publication in MNRAS that reaches similar conclusions.  We would
  like to thank Andrea Merloni, Mike Nowak, Rob Fender, and Rashid
  Sunyaev for helpful discussions and the anonymous referee for
  several important suggestions on how to improve this paper.  Support
  for this work was provided by the National Aeronautics and Space
  Administration through Chandra Postdoctoral Fellowship Award Number
  PF3-40026 issued by the Chandra X-ray Observatory Center, which is
  operated by the Smithsonian Astrophysical Observatory for and on
  behalf of the National Aeronautics Space Administration under
  contract NAS8-39073.}

\end{document}